\documentclass[aps,prd,twocolumn,showpacs,preprintnumbers,nofootinbib]{revtex4}
\usepackage{amssymb,latexsym}
\usepackage{amsmath,amsbsy}
\usepackage{epsfig}
\DeclareMathOperator{\Pperp}{\mathbf{P}^{\perp}}

\DeclareMathOperator{\kperp}{\mathbf{k}^{\perp}}

\DeclareMathOperator{\dperp}{\mathbf{\Delta}^\perp}

\begin{document}

\preprint{NT-UW 02-19}

\title{Generalized parton distributions for $q\bar{q}$ pions}

\author{B.~C.~Tiburzi}
\author{G.~A.~Miller}

\affiliation{Department of Physics  
        University of Washington      
        Box 351560
        Seattle, WA 98195-1560}

\date{\today}

\begin{abstract}
We use a two-body, light-cone  pion wave function
(which vanishes when one of the constituents carries all of the plus-momentum)
to construct a double distribution. 
We show that the resulting generalized parton distribution
has incorrect behavior at the crossover between kinematic regions. 
Furthermore the positivity constraints are not respected.
On the other hand, the field-theoretic, two-body double distribution 
gives  generalized parton distributions that have physically correct crossover behavior
and respect the positivity constraints independent of the model wave function
used. However, the sum rule and polynomiality conditions cannot be satisfied
with a two-body wave function alone since Lorentz symmetry is not maintained.
\end{abstract}


\maketitle

\section{Introduction}
In recent years generalized parton distributions (GPDs) \cite{Muller:1998fv,Ji:1996ek,Radyushkin:1996nd} have generated a considerable amount of attention. 
These distributions stem from hadronic matrix elements that are both non-diagonal with respect to hadron states and involve quark and gluon operators 
separated by a light-like distance. Thus physics of both inclusive (parton distributions, e.g.) and exclusive (form factors, e.g.) reactions is contained 
in the GPDs. At the leading-twist level, these new structure functions describe the soft physics of a variety of hard exclusive processes (see the 
reviews \cite{Ji:1998pc}). 

Recently two-body light-front wave function models of the pion have been used to obtain
GPDs \cite{Mukherjee:2002gb}
based on  double distributions (DDs)
\cite{Radyushkin:1997ki,Radyushkin:1998es}.
Such wave functions, based on solving the Weinberg
equation \cite{Frederico:2001qy} and carrying the total charge of the pion
vanish when either the quark or anti-quark carries all of the plus-momentum of
the pion. The purpose of this note is to show that the use of such a model
leads to  GPDs with undesirable features.  
Therefore we believe that constructing simple models with the relevant physics that
satisfies the reduction relations for GPDs remains a considerable challenge. 

We proceed using the following logic. 
The beginning,  in section \ref{pion},  discusses 
the pion model GPD
introduced in \cite{Mukherjee:2002gb} by  reviewing
the power-law wave function used 
and showing how the DD was obtained from the expression for  the form factor.
We also indicate the reduction relations required of the GPDs.
In the following section 
(\ref{petit}), the physics at the crossover $(X = \zeta)$
between kinematic regions  is reviewed in terms of the Fock space
representation.
Internal consistency
demands a model with a vanishing distribution function (at zero plus-momentum)
to have vanishing GPDs at the crossover. The proposed model is shown not to 
satisfy this restriction. Furthermore, the
realistically modified model of Ref.~\cite{Mukherjee:2002gb}
does not match up to the physical intuition afforded by the Fock space 
representation which relates small $x$ behavior of $q(x)$ to GPDs at the crossover.
Section \ref{poscon} investigates whether the models respect the positivity constraints. The realistically augmented model satisfies 
positivity, while its underlying wave-function-based model violates the
constraint for all $X>\zeta$. The Fock-space representation is then used to obtain 
DDs  by using the valence wave function  consistently in  the 
relevant matrix element
(section \ref{DDfock}). These  DDs
 are shown to violate Lorentz symmetry by retaining $\zeta$-dependence.
The  two-body model GPD obtained from these DDs
vanishes at the crossover  and satisfies the positivity constraints. 
As a result of the $\zeta$-dependence,
however, moments of these GPDs do not satisfy the polynomiality conditions. 
In particular the sum rule cannot be satisfied in the valence sector alone. 
Additionally in the Appendix, we scrutinize the covariant scalar model used in 
the Appendix of \cite{Mukherjee:2002gb} to justify the construction of the DD. 
Parametric differentiation of a covariant field-theoretic amplitude 
is shown not to preserve positivity. A brief summary is presented in section \ref{sumy}.

\section{Pion model wave function} \label{pion}
Here we review the basics of the model proposed by \cite{Mukherjee:2002gb} starting from the two-body wave function and ending at the double distribution. 
Motivated by a numerical solution to an effective two-body equation for the pion \cite{Frederico:2001qy}, the following wave function is adopted
\begin{equation} \label{psi}
\psi(x, \kperp) = \frac{N}{\sqrt{x(1-x)} \big[ a(x) + b(x) \frac{\kperp^2}{m^2} \big]^2},
\end{equation}
where $a(x)$ and $b(x)$ are the dimensionless functions
\begin{equation}
a(x) = 1 + \frac{s^2 (x - \frac{1}{2})^2}{x (1-x)}, \;\;\; b(x) = \frac{s^2}{4 x(1-x)},
\end{equation}
with $s$ a dimensionless parameter of the model and $N$ the normalization constant. One then determines the normalization $N$ from the condition
\begin{equation} \label{norm}
\int dx d\kperp \big|\psi(x, \kperp)\big|^2 = 1, 
\end{equation}
which makes us think of $\psi$ as a wave function that in some way effectively represents contributions from higher Fock components. 

Having modeled the physics in the lowest ($q \bar{q}$ symmetric)  Fock component, the Drell-Yan formula \cite{Drell:1969km} is used for the form factor 
\begin{equation} \label{DY}
F(t) = \int dx d\kperp \psi^*\big(x, \kperp + (1 - x)\dperp\big) \psi\big(x, \kperp\big),
\end{equation}
where $t = - \dperp^2$. Using the explicit structure of the wave function 
Eq. \eqref{psi}, the form factor can be manipulated into the form
\begin{equation}
F(t) = \int_0^1 dx \int_0^1 dy\;  F(x,y;t),\label{manip}
\end{equation}
with the DD
\begin{equation} \label{DD}
F(x, y; t) = \theta(1 - x - y) \frac{6 y ( 1 - x - y)/(1-x)^3}{A \big[a(x) - \frac{t}{m^2} y (1 - x - y) b(x) \big]^3}.
\end{equation}
Here $A$ is just a constant related to the normalization by $A = \frac{3}{4\pi} 
\frac{s^2}{m^2} \frac{1}{N^2}$. This DD has the proper support \cite{Radyushkin:1996nd} and is \emph{M\"unchen}
symmetric \cite{Mankiewicz:1997uy}: $F(x, y;t) = F(x, 1 - x- y; t)$. In the forward limit, the $y$-dependence factorizes 
\begin{equation} \label{fact}
F(x,y;0) = \theta(1 - x -  y) h(x,y) q(x) 
\end{equation}
into the normalized profile function $h(x,y)$
\begin{equation}
h(x,y) = \frac{6 y (1-x-y)}{(1-x)^3}
\end{equation}
satisfying
\begin{equation} \label{prof}
\int_0^{1-x}dy\;  h(x,y) = 1.  
\end{equation}
The profile function $h(x,y)$ is identical to the asymptotic quark distribution amplitude model suggested in \cite{Radyushkin:1998es}. 
The properties Eqs. \eqref{fact} and \eqref{prof}
are essential so that the total charge $F(0)$ looks like an integral of the quark distribution function $q(x)$, namely
\begin{equation} \label{q}
q(x) = \int d\kperp \big|\psi(x,\kperp)\big|^2 = \frac{1}{A a(x)^3}.
\end{equation}
Notice that $q(0)=0$. 

The ingenuity of DDs comes about when we construct the GPD \cite{Radyushkin:1997ki} via
\begin{equation} \label{gpd}
\mathcal{F}_{\zeta}(X, t) = \int_0^1 dx \int_0^1 dy \; F(x,y;t) \delta(X - x - \zeta y). 
\end{equation}
In this form,  the sum rule for the form factor
\begin{equation} \label{sum}
\int_0^1 dX \; \mathcal{F}_{\zeta}(X,t) = F(t)
\end{equation}
and  the polynomiality constraints on moments of $\mathcal{F}$ are immediately satisfied. 

In summary, the above model
satisfies the spectral properties and
reduction relations required of GPDs. 
Moreover, unlike previously used Gaussian \emph{Ans\"atze} for $\zeta = 0$ GPDs, there is non-trivial interplay 
between $X$, $\zeta$ and $t$ dependence.

\section{At the crossover} \label{petit}
The first inconsistency of the $q\bar q$ pion
model comes from behavior at the crossover, i.e.~when $X = \zeta$. Using the Fock space representation of 
deeply virtual Compton scattering \cite{Brodsky:2001xy,Diehl:2000xz},
we first derive a result 
that focuses on the small-x behavior of the model wave function. 

Since $q(0) = 0$ above, let us consider models such that the quark distribution function $q(x)$ vanishes at $x = 0$. 
This quark distribution function, however, is a model of all the higher Fock components. Although we cannot generally disentangle these components from a given model for $\psi$, we implicitly have the restriction
\begin{equation} \label{qzero}
0 = q(0) = \sum_{n} \sum_{j} \int_{\{n\}} \sum_{\lambda_i} \delta(x_{j}) \; \big|\psi_{n}(x_{i},\mathbf{k}^\perp_{i},\lambda_i)\big|^2,  
\end{equation}
where the sum runs over all $j$-quarks in each $n$-body Fock state and the integration measure is given by
\begin{equation}
\int_{\{n\}} = \int \prod_{i = 1}^{n} \frac{dx_{i} d\mathbf{k}^\perp_{i}}{16 \pi^3} 16 \pi^3 \delta\Big(1 - \sum_{i=1}^{n} x_{i}\Big) \delta\Big(\sum_{i=1}^{n} \mathbf{k}^\perp_{i}\Big).
\end{equation} 
But since each term's integrand in Eq. \eqref{qzero} is positive definite, we must have
\begin{equation} \label{biggie}
\psi_{n}(\ldots, x_{j} = 0, \ldots) = 0
\end{equation}
whenever $j$ corresponds to a quark. 

This strong result (which applies {\em only} to models with $q(0)=0$)
has a direct implication for GPDs at the crossover. Approaching the crossover from above we have
\begin{widetext}
\begin{equation} \label{crossfock}
\mathcal{F}_{\zeta}(\zeta,t) \propto \sum_{n} (1 - \zeta)^{1 - \frac{n}{2}} \sum_{j} 
\int_{\{n\}} \sum_{\lambda_i} \delta(\zeta - x_{j}) \psi_{n}^*(x^\prime_i, \mathbf{k}^\prime_i{}^\perp,\lambda_i) 
\psi_{n}(x_{i},\mathbf{k}^\perp_i,\lambda_i),
\end{equation}
\end{widetext}
with the primed variables given by
\begin{align}
x^\prime_{j} & = 0 \quad \quad \quad \quad \mathbf{k}^\prime_j{}^\perp = \mathbf{k}^\perp_j - \dperp \\
x^\prime_{i \neq j} & = \frac{x_i}{1 - \zeta} \quad \quad \mathbf{k}^{\prime\perp}_{i \neq j} = \mathbf{k}^\perp_j + x^\prime_i \dperp. 
\end{align}
Thus since the struck quark rebounds with zero plus-momentum, Eq. \eqref{biggie}
forces $\mathcal{F}_{\zeta}(\zeta,t)$ to vanish.\footnote{There is the possibility of the \emph{Gibb's phenomenon}, for which the sum of infinitely 
many zeros in Eq.~\eqref{crossfock} need not vanish. This is not expected; Fock components are not independent.} Consistency requires any effective modeling of the higher Fock components with a vanishing quark distribution
at zero plus-momentum to mandate GPDs to vanish at the crossover. Here we have considered the GPD for a scalar particle. For a spin one-half bound state, the correct generalization of the above result is that the imaginary part of the deeply virtual Compton amplitude $\mathcal{M}$ vanishes. This is because the generalized parton 
distributions $H(X,\zeta,t)$ and $E(X, \zeta,t)$ enter as the linear combination
\begin{equation}
\Im \mathcal{M} \propto \frac{\sqrt{1 - \zeta}}{1 - \frac{\zeta}{2}} H(\zeta,\zeta,t) - \frac{\zeta^2}{4 ( 1 - \frac{\zeta}{2}) \sqrt{1 - \zeta}} E(\zeta, \zeta, t),
\end{equation}
which has a light-cone Fock space decomposition similar to Eq. \eqref{crossfock}; the only differences are labels for identical initial and final bound-state spin.  

The pion model of section \ref{pion} has a quark distribution function  \eqref{q} which vanishes at zero plus-momentum. Thus the model effectively takes each Fock component to vanish at $x = 0$, and hence we would expect this model's GPD $\mathcal{F}_\zeta(X, t)$ to vanish at the crossover. Using Eq. \eqref{gpd} for the model DD \eqref{DD}, we find it is not the case (as one could already see from Figure $4$ in \cite{Mukherjee:2002gb})
\begin{equation} \label{cross}
\mathcal{F}_\zeta(\zeta,t) = \int_{0}^{1} dy \; F\big(\zeta(1 - y),y;t\big). 
\end{equation}

\begin{figure}
\begin{center}
\epsfig{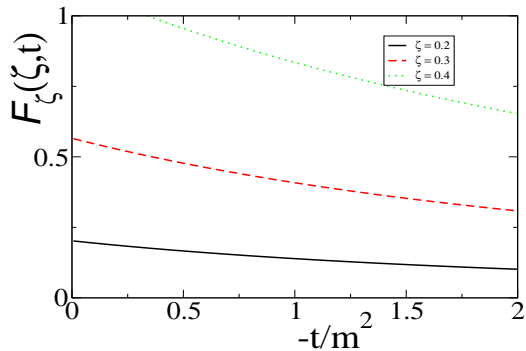}
\caption{Value of the generalized parton distribution at the crossover: $\mathcal{F}_{\zeta}(\zeta,t)$. Here we choose the various values for $\zeta$  and plot the crossover value as a function of $\frac{-t}{m^2}$ for the model parameter $s = 0.95$.}
\label{fone}
\end{center}
\end{figure} 

In Figure \ref{fone}, we show the value at the crossover as a function of $\frac{-t}{m^2}$ for a few values of $\zeta$. 
Despite the vanishing of the quark distribution  function at zero $x$, the model GPD of \cite{Mukherjee:2002gb}
does not vanish at the crossover. Said another way, the non-vanishing 
behavior at the crossover is indicative of higher Fock space contributions which are not present in the quark distribution function, 
i.e.~the constituent quarks in Eq. \eqref{psi} have no sub-structure. The model GPDs appear to have 
more physics than the underlying wave function.
 
\begin{figure}
\begin{center}
\epsfig{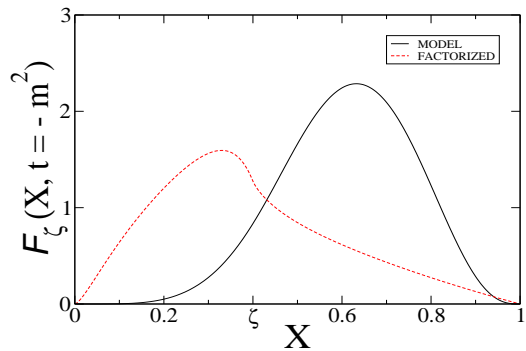}
\caption{Comparison of the model GPD Eq. \eqref{gpd} (MODEL) with the realistic, factorized form Eq. \eqref{gpdreal} (FACTORIZED)
for fixed $\zeta = 0.4$ and $t = - m^2$. The model parameter is again chosen to be $s = 0.95$. 
Physically we expect the original model to vanish at $X = \zeta$ while the realistic version should be greatly enhanced there. }
\label{ftwo}
\end{center}
\end{figure}

Acknowledging the limits of the model quark distribution function \eqref{q}, the authors \cite{Mukherjee:2002gb} suggest
implanting a more realistic distribution
\begin{equation} \label{qreal}
q^{R}(x) = \frac{3}{4} \; \frac{ 1 - x}{\sqrt{x}} 
\end{equation}
and modify the DD accordingly by using a factorization assumption
\begin{equation} \label{gpdreal}
F^{R}(x,y;t) = F(x,y;t) \frac{q^{R}(x)}{q(x)}.
\end{equation}
Modifying the quark distribution function according to Eq. \eqref{qreal} must change the
wave function. Considering the singular behavior, the effective wave function $\psi_{\text{eff}} = \psi \sqrt{\frac{q^R}{q}}$ 
must be dramatically enhanced at small-$x$. 
Thus one consequence we expect is a sizable enhancement to the GPD at the crossover due to the form of Eq. \eqref{crossfock}. 
Comparing the original GPD \eqref{gpd} to the realistic one calculated from the
DD in Eq. \eqref{gpdreal} in Figure \ref{ftwo}, we notice that the integral of
Eq.~(\ref{cross}) leads to 
only a small enhancement. Thus the factorized assumption does not support an overlap interpretation in terms of the effective wave function. 
 
One could ask why bother looking at the crossover in the first place. Experimentally \cite{Saull:1999kt} the DVCS amplitude is accessed
from the dominant Bethe-Heitler contamination through beam asymmetries. At first the interfering QED radiative processes
were seen as a hindrance to unmasking the DVCS contribution to the total cross-section. Now however, they seem more of a virtue, allowing 
access to both the real and imaginary parts of the virtual Compton amplitude. The beam-spin asymmetry \cite{Diehl:1997bu},
for example, allows the imaginary part of the amplitude to be measured. At leading twist, the imaginary part
involves the GPDs evaluated at the crossover $X = \zeta$. From the point of view of modeling, 
it would be nice to understand measurements of the beam-spin asymmetry. The
above model, however has been constructed \cite{Frederico:2001qy}
so that it should yield a vanishing value at the crossover. Any  non-zero result
of a $q\bar q$  model, is simply an artifact of Eq.~(\ref{manip}). As we shall
show below in Sect.~\ref{DDfock}, the integrand of that expression is not the
DD because it is not obtained
from is definition as a specific matrix element of field operators.
A better determined $q\bar q$ 
wave function will not lead to a better understanding of GPDs at the crossover.

On the other hand, if one is concerned with the real part of the virtual Compton amplitude (which enters in the charge asymmetry
\cite{Belitsky:2000gz}), the GPDs appear in a (principle value) integral 
weighted by the hard scattering amplitude. The crossover effects are still important 
because a factor of $\frac{1}{X - \zeta}$ is present and the GPDs are likely to have discontinuous derivatives. 
The only sophisticated model study on the market \cite{Petrov:1998kf} (which arguably does QCD in the large $N_{c}$ limit) 
shows considerable contribution ($\sim 60 \%$) to the cross-section from near the crossover. Lastly one could hope to measure the GPDs directly
via double DVCS \cite{Guidal:2002kt}.

\section{Positivity Constraints} \label{poscon}

The result proved in the light-cone Fock representation in section \ref{petit}
concerning behavior at the crossover stemming from the small-$x$ limit of the  quark distribution also follows from the positivity constraints for GPDs. 
Originally these constraints appeared in \cite{Radyushkin:1998es,Pire:1998nw} and were derived from the positivity of the density matrix by restricting 
the final-state parton to have positive plus-momentum (and ignoring the contribution from $E(X,\zeta,t)$ for the spin-$\frac{1}{2}$ case).  Although the 
matrix elements involved for GPDs are off diagonal, they are still restricted by positivity and their diagonal elements. Correcting the constraints
for the presence of the $E$-distribution was first done in \cite{Diehl:2000xz}. 
By considering the positivity of the norm on Hilbert space, stricter constraints for the spin-$\frac{1}{2}$ distributions $H$ and $E$ have recently 
appeared as well as constraints for the full set of twist-$2$ GPDs \cite{Pobylitsa:2001nt}. 

\begin{figure}
\begin{center}
\epsfig{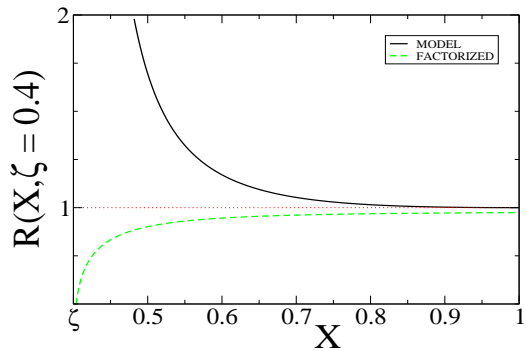}
\caption{Comparison of the model GPD Eq. \eqref{gpd} (MODEL) with the realistic, factorized form Eq. \eqref{gpdreal} (FACTORIZED)
for fixed $\zeta = 0.4$ at $t = 0$. The model parameter is again chosen to be $s = 0.95$. Here we plot the ratio $R(X,\zeta)$
appearing in equation \eqref{posi} as a function of $X > \zeta$. Positivity constrains this ratio to be less than one.}
\label{fthree}
\end{center}
\end{figure}

For the scalar pion case there is of course no contribution from the non-existent $E$-distribution and hence the original bounds are actually correct. 
When $X \geq \zeta$ the positivity constraint for the spin-$0$ pion reads
\begin{equation} \label{posi}
R(X,\zeta) \equiv (1 - \zeta/2) \frac{ \big|\mathcal{F}_{\zeta}(X,0)\big| }{ \sqrt{q \big(X\big) q\big(\frac{X - \zeta}{1 - \zeta}\big) } }   \leq 1.
\end{equation}
Of course the result holds for finite $-t$, however since the function
$\mathcal{F}$
decreases with $-t$, so that \eqref{posi} is the tightest constraint.
Notice the bound at $X = \zeta$ corresponds to the result derived in section \ref{petit} provided of course the hypothesis is met: $q(0) = 0$. 
The GPD derived from Eq. \eqref{DD} violates the constraint for all $X > \zeta$ as depicted in Figure \ref{fthree} for $\zeta = 0.4$, while 
the augmented model \eqref{gpdreal} satisfies the constraint.   

In the realistic quark distribution model \eqref{qreal} the positivity constraint is quite lenient as $q^R (0) = \infty$ allows for no bound on the 
GPD at $X = \zeta$ in equation \eqref{posi}.  Consequently near $X = \zeta$ the GPD is only very loosely constrained. 
In general, a small-$x$ factorized fix-up of $q(x)$ will not guarantee that the positivity constraint will be met.

\section{Double distributions in Fock space} \label{DDfock}

As we have seen the pion model proposed in \cite{Mukherjee:2002gb} violates the positivity constraint near the crossover. 
The realistic fix-up, while satisfying positivity, does not agree with the intuition provided by the light-cone Fock space representation 
which, e.g., connects small-$x$ behavior to GPDs at the crossover. 
Our solution to this dilemma is to define the double distribution consistently using the same Fock space representation implicitly 
used in Eqs. \eqref{q} and \eqref{DY}. To achieve this, we truncate
the Fock space at the valence sector as we did previously \cite{Tiburzi:2001ta}.

In the two-body truncated Fock space the pion bound state of momentum $P$ appears as
\begin{equation} \label{twofock}
| \pi(P) \; \rangle  = \int \frac{dx d\kperp}{\sqrt{16 \pi^3 x ( 1-x)}}  \psi(1,2) \;
b^\dagger(1) \; d^\dagger\big(2) \; | \; 0 \rangle ,
\end{equation}
where $1$ corresponds to the momentum $xP^+, \kperp$, while $2$ represents $(1-x)P^+, \mathbf{P}^\perp - \kperp$.  
Without loss of generality, we have chosen particle $1$ to be the quark and $2$ the anti-quark. One can use
the theory's Galilean invariance to write $\psi(1,2) = \psi(x, \kperp - x \mathbf{P}^\perp)$, which is the same as  $\psi(1-x, x \mathbf{P}^\perp - \kperp)$
by $q\bar{q}$ symmetry.
We omit helicity labels to make our analysis parallel to that of \cite{Mukherjee:2002gb}.
The bound state is normalized according to 
\begin{equation}
\langle \; \pi(P^\prime)\; | \; \pi(P)\; \rangle  = 16 \pi^3 P^+ \delta(P^{\prime +} - P^+) 
\delta(\mathbf{P}^{\prime \perp} - \Pperp),
\end{equation}
which in turn implies the following normalization for the valence wave function
\begin{equation} \notag
\int dx d\kperp \big| \psi(x,\kperp) \big|^2 = 1. 
\end{equation}
As above, we can think of Eq.~(\ref{norm}) as the consequence of truncating the Fock space to only the valence sector, i.e.~$\psi$ must be an effective
two-body wave function. The constituent quarks in our bound state, however, have no sub-structure nor should we gain Lorentz invariance from 
Eq. \eqref{norm}. 

The good component of the quark field $\psi_+ = \mathcal{P}_+ \psi$ (with $\mathcal{P}_{+} = \gamma^- \gamma^+ /2$) has a  mode expansion 
\begin{widetext}
\begin{equation}
\psi_+(z^-, \mathbf{0}^\perp) = \int \frac{dk^+ d\kperp \theta(k^+)}{16\pi^3 k^+} \sum_{\lambda} \Big[ b(k^+,\kperp) \mathcal{P}_+ u_\lambda(k^+,\kperp)
e^{-i k^+ z^-} + d^\dagger(k^+,\kperp) \mathcal{P}_+ v_\lambda(k^+,\kperp) e^{i k^+ z^-}  \Big],
\end{equation} 
\end{widetext}
where the mode operators satisfy the anti-commutation relation
\begin{equation}
\Big\{ b(k^+,\kperp), b^\dagger(p^+,\mathbf{p}^\perp)  \Big\} = 16 \pi^3 k^+ \delta(k^+ - p^+) \delta(\kperp - \mathbf{p}^\perp)
\end{equation}
and similarly for $d$ and $d^\dagger$.

These equations are all consistent with the starting point of the pion model in section \ref{pion}. For instance:
the quark distribution function is identical to Eq. \eqref{q}
\begin{equation}
q(x)  \equiv \frac{1}{2} \int \frac{dz^-}{2\pi} e^{i x P^+ z^-} \langle \pi(P) \; | \bar{\psi}(0) \gamma^+ \psi(z^-) | \; \pi(P) \rangle. 
\end{equation}
Additionally the matrix element definition of the form factor lines up with the Drell-Yan formula
\begin{equation}
F(t) \equiv \frac{1}{2 P^+} \langle \pi(P^\prime)\;  | \bar{\psi}(0) \gamma^+ \psi(0) | \; \pi(P) \rangle,
\end{equation}
which is given by Eq. \eqref{DY} with $P^\prime = P + \Delta$ and $ t = \Delta^2 = - \mathbf{\Delta}^\perp{}^2$, i.e.~$\Delta^+ = 0$.  

\begin{widetext}
\subsection{Derivation of the DD}
The DD is defined from the non-diagonal matrix element of bilocal field operators \cite{Radyushkin:1998es,Radyushkin:1997ki}. 
To derive the DD consistently, we should respect this underlying field theoretic construction: 
\begin{equation} \label{DDmel}
\frac{1}{2 P^+ (1 - \zeta/2)} \langle \pi(P^\prime) \; | \bar{\psi}(0) \gamma^+ \psi(z^-)  | \; \pi(P) \rangle = \int_0^1 \int_0^1 
\theta(1 - x - y) F(x,y;t) \Big[e^{-i(xP^+ - y \Delta^+)z^-} - e^{i\big(x P^+ + (1-y) \Delta^+\big) z^-}   \Big] dx dy.
\end{equation}

Using the truncated Fock space expansion of the bound state \eqref{twofock}, we can express the matrix element as
\begin{multline} \label{midstep}
\langle \pi(P + \Delta) \; | \; \bar\psi(0) \gamma^+ \psi(z^-)  \; | \; \pi(P) \rangle  = 
2 \int dk^+ d\kperp \theta(k^+ + \Delta^+) \theta(k^+) \\ 
\times \psi^*(Y,\kperp + (1-Y) \dperp) \psi(X, \kperp) \Big[ e^{- i k^+ z^-} - e^{i (k^+ + \Delta^+)z^-} \Big],
\end{multline}
where we have used the temporary replacements: $X = k^+/P^+$ and $Y = \frac{k^+ + \Delta^+}{P^+ + \Delta^+}$.
To find the DD, we merely need to cast the above into the form dictated by 
Eq. \eqref{DDmel}. Thus it remains to introduce $x,y$. Before doing so explicitly, it is wise to make a few comments. 
Given the $z^-$ dependence of Eq. \eqref{midstep}, we must introduce $x, y$ only via $k^+ = x P^+ - y \Delta^+$. In turn, however, 
the resulting $F(x,y;t)$ must depend on $\Delta^+ = - \zeta P^+$. This is obvious looking at the two-body partonic cartoon 
shown in Figure \ref{ffour} for the DD. In fact, one can see that any $N$-body 
Fock component contribution to the DD will retain $\zeta$ dependence. This should not be surprising: the light-cone wave functions are not 
covariant objects. It is only when summing over all Fock component overlaps that the $\zeta$ dependence in the DD cancels. 

 \begin{figure}
\begin{center}
\epsfig{file=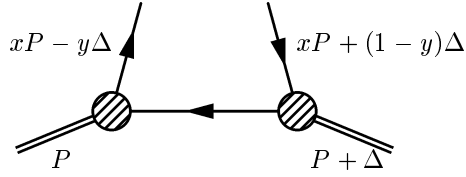}
\caption{Partonic diagram for valence two-body double distribution.} 
\label{ffour}
\end{center}
\end{figure}

To cast Eq. \eqref{midstep} into a form from which we can read off the DD, we must use an explicit form for the wave function. 
Using the model wave function \eqref{psi} above, we can write
\begin{align}
I & \equiv \frac{1}{m^2 N^2} \sqrt{X(1-X)Y(1-Y)} \int d\kperp \psi^*(Y,\kperp - Y \dperp) \psi(X,\kperp) \notag \\
  & = \int \frac{d\kperp}{\big[ a(X) + b(X) \kperp^2 \big]^2    \big[ a(Y) + b(Y) \big(\kperp + (1-Y)\dperp^2\big)\big]^2}.     
\end{align}
Notice that now $\kperp$ and $\dperp$ have been scaled by $m$ to be dimensionless.
Exponentiation of the denominators via $A^{-2} = \int_{0}^{\infty} \alpha e^{-\alpha A} d\alpha$ allows for a Gaussian integral over 
$\kperp$ to be performed. The result is
\begin{equation}
I = \pi \int_{0}^{\infty} \int_{0}^\infty d\alpha_1 d\alpha_2 \frac{\alpha_1 \alpha_2}{\alpha_1 b(X) + \alpha_2 b(Y)} \exp\Bigg\{-\alpha_1 a(X) - \alpha_2 a(Y)
-\frac{\alpha_1 \alpha_2 b(X) b(Y) (1-Y)^2 \dperp^2}{\alpha_1 b(X) + \alpha_2 b(Y)}\Bigg\}.
\end{equation} 
Since $b \geq 0$, we define the new variables $\lambda$,$\beta$: $\lambda = \alpha_1 b(X) + \alpha_2 b(Y)$ and $\alpha_1 b(X) = \lambda \beta$. Performing the resultant integral over $\lambda$ yields 

\begin{equation}
I = \frac{1}{[b(X)b(Y)]^2} \int_{0}^{1} \frac{\beta (1-\beta) d\beta}{\Big[    \beta \frac{a(X)}{b(X)} + (1 - \beta) \frac{a(Y)}{b(Y)} + \beta(1-\beta) (1-Y)^2 \dperp^2\Big]^3}.
\end{equation}
Next we introduce $y$ and $\alpha$ by $y = (1 - \alpha^2) \beta$ and $X = \alpha^2 + \zeta y$. Under this change of variables 
$\theta(\beta) \theta(1-\beta) = \theta(y) \theta(1-y) \theta(1-\alpha^2) \theta(1 - \alpha^2 - y)$ and further 
$d\beta dk^+ = 2 P^+ |\alpha| \frac{|1 - \alpha^2 - \zeta y| }{(1-\alpha^2)^2} dy d\alpha$. The resulting integral is even in $\alpha$ 
and hence we introduce $x = \alpha^2$ so that $\int_{-1}^{1} d\alpha |\alpha| \ldots = \int_0^1 dx \ldots$. Noticing that 
$\theta(1 - \alpha^2)$ = $\theta(x)\theta(1-x)$, we can write out the DD from Eq. \ref{midstep}
\begin{equation} \label{DDmess}
F(x,y;t)  = \frac{6 s^2 (1-X)\theta(X) \theta(1 - X) \theta(X - \zeta)}{4 A (1 - \zeta/2)
  \sqrt{X(1-X) Y(1-Y)}\; b^2(X)b^2(Y)} \frac{y(1-x-y)/(1-x)^4}{\Big[
  \frac{y}{1-x} \frac{a(X)}{b(X)} + \frac{1 - x- y}{1 - x} \frac{a(Y)}{b(Y)}  +
  \frac{y(1-x-y)}{(1-x)^2} (1-Y)^2 \dperp^2 \Big]^3} 
\end{equation} 
where now the labels are $X = x + \zeta y$ and $Y = \frac{x- (1-y) \zeta}{1 - \zeta}$. This DD based on the field-operator matrix element \eqref{DDmel} is 
clearly different from Eq. \eqref{DD}. 
\end{widetext}

\subsection{Properties} 
The use of  a $q\bar q$ wave function is a violation of Lorentz symmetry which
has drastic  consequences for the computed DD of Eq.~(\ref{DDmess}). The
violation of
Lorentz symmetry and the  \emph{M\"unchen} symmetry is manifest in 
the $\zeta$-dependence of  $F$ in Eq. \eqref{DDmess}. This $\zeta$-dependence spoils the sum rule for the electromagnetic form factor. Additionally the DD is restricted by $\zeta$ in $x-y$ space as a result of the $\theta$-functions that maintain the light-cone spectrum condition.  One way to obtain the full support properties of the DD is to use higher Fock components. Alternatively, one could imagine constructing a $\zeta$-dependent function for the excluded region of phase space which has the properties necessary to cancel the $\zeta$-dependence of $F$ above. In essence, this would be an attempt at solving for the (light-cone) instantaneous kernel for $\psi$ and utilizing crossing to access all kinematical regions \cite{Tiburzi:2001je}. One can do so in the case of an instantaneous interaction because there are no true higher Fock states. 

Taking the $\zeta \to 0$ limit of Eq. \eqref{DDmess} results in the pion model DD above \eqref{DD}. 
Thus we additionally have the reduction relation at $t = 0$
\eqref{fact}. It is not surprising that the $\Delta^+=0$ limit corresponds to the model of \cite{Mukherjee:2002gb}. They extract the DD from the 
Drell-Yan formula which translates to calculating the matrix element in $\eqref{DDmel}$ using the $\Delta^+ = 0$ frame. It seems reasonable that 
calculation of the Lorentz invariant DD can be done in the $\Delta^+ = 0$ frame; however, the result does not satisfy positivity.

\begin{figure}
\begin{center}
\epsfig{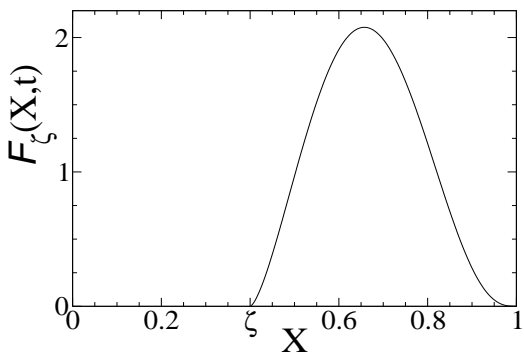}
\caption{Plot of the positivity constrained GPD Eq. \eqref{focky} for $\zeta = 0.4$ and with the model
parameter $s = 0.95$. We choose $\dperp^2 = 0.6 m^2$ so that $t$ will be essentially the same as in Figure \ref{ffour}.}
\label{ffive}
\end{center}
\end{figure}

Using Eq. \eqref{cross}, we can show that the GPD resulting from Eq. \eqref{DDmess} vanishes at the crossover. This follows simply from the $Y \to
0$ behavior of $F$. Working with wave functions directly illuminates this. 
From Eq. \eqref{midstep}, we can write down the GPD
\begin{widetext}
\begin{multline} \label{focky}
\mathcal{F}_\zeta(X,t) =  \frac{1}{1-\zeta/2} 
\int d\kperp \Bigg[ \theta(X) \theta(1-X) \theta(X - \zeta) \psi^*(Y, \kperp + (1 - Y) \dperp)  \psi(X,\kperp)\\ 
- \theta(-X) \theta(X - \zeta + 1) \theta (\zeta - X) \psi^*\big(-X/(1-\zeta), \kperp + (1 + X/(1-\zeta) \dperp \big) \psi(\zeta - X, \kperp)      
\Bigg].
\end{multline} 
\end{widetext}
From the above expression, we directly see the GPD's vanishing at the crossover
results from the small-$x$ behavior of the model wave function. Furthermore,
as is generally true \cite{Diehl:2000xz}, direct application of the Schwartz inequality yields the positivity constraint \eqref{posi}. 
In Figure \ref{ffive}, we plot the GPD Eq. \eqref{focky} which can also be obtained from \eqref{DDfock} by means of Eq. \eqref{gpd}. There 
is a big difference between Figures \ref{ffour} and \ref{ffive}.

Lastly one could try to incorporate a realistic quark distribution by using a 
factorized \emph{Ansatz} for the double distribution. Implementing Eq. \eqref{gpdreal} for the DD above \eqref{DDmess}, however, does not affect the 
vanishing of the GPD at  the crossover. This is again contrary to our observation from Eq. \eqref{crossfock} and hence the factorized assumption must 
not change the small-$x$ behavior of the wave function. A more physical \emph{Ansatz} is effected by directly altering the behavior 
at the end-point: $\psi \to \psi(x,\kperp) \sqrt{\frac{q_R(x)}{q(x)}}$. This alteration maintains the positivity constraint regardless of the 
functional form of $q_R(x)$. The drawbacks of this form are clear: resulting GPDs are discontinuous and the particular form of Eq. \eqref{qreal} leads 
to a singular GPD at the crossover.  The effect of $q_R(x)$ is to mimic contributions from higher Fock states. Continuity of the GPD arises from 
relations between Fock components at vanishing plus-momentum.  A perturbative model confirms this \cite{Tiburzi:2002mn}. Constructing simple light-cone 
wave function models for GPDs consistent with the reduction relations and positivity remains a challenge.

\section{Summary} \label{sumy}
The DD based model of GPDs proposed in \cite{Mukherjee:2002gb} is appealingly
ingenious, but has been shown to have undesirable properties. Behavior at the
crossover is inconsistent with intuition.
Positivity is violated because additional time-orderings involving non-two-body
contributions are omitted. Alternately we can view violation of the positivity
constraint as resulting from parametric differentiation of a covariant
field-theoretic amplitude. As shown in the Appendix, such differentiation 
does not preserve positivity.
On the other hand, the light-front Fock representation of GPDs is intuitively physical, satisfies positivity, 
but manifestly breaks Lorentz symmetry.  

We present these problems  for two reasons.
Firstly, to treat other theoretical attempts in this direction with
caution. For example,
one could imagine applying the procedure in \cite{Mukherjee:2002gb} to a better determined (or better tuned)
valence wave function. We have shown that this will not result in a more accurate determination of the GPD.
The procedure is not internally consistent (i.e.~the positivity constraints will be violated) and one must 
ultimately augment the wave function with a
realistic quark distribution using an untested factorization \emph{Ansatz}.
This augmentation is an attempt to
alter the small-$x$ behavior of the wave function.  However, the resulting GPDs do not match 
up with the physical intuition provided by the light-cone Fock representation
at the crossover which closely relates the low-$x$ quark distribution functions
to GPDs at the crossover.    
Secondly we believe  this cautionary note could be  important to the
experimentalists
hoping to interpret future data. 
The reverse of this procedure seems
to suggest one can learn about the effective valence wave function from experimental access to GPDs. 
There is no such correspondence: the model GPD from Eq. \eqref{DD} appears to
contain more
physics than the valence wave function from which it was derived.

Lastly we remark that the proposed model \cite{Mukherjee:2002gb}
does provide an alternative to Gaussian based forms---allowing for 
the correct $t$-dependence. While the model is useful as an 
analytical result, one cannot  take the $X$ and $\zeta$ dependence seriously. 

\begin{acknowledgments}
We are indebted to A.~V.~Radyushkin for discussion and correspondence. 
This work was funded by the U.~S.~Department of Energy, grant: DE-FG$03-97$ER$41014$.  
\end{acknowledgments}

\appendix

\section{Double distributions in a scalar model}
Above we have seen that the wave function based DD \eqref{DD} does not satisfy the positivity constraint
which creates doubt about the uniqueness of its construction. 
In their Appendix, the authors \cite{Mukherjee:2002gb}, however, do not treat the wave function as in section \ref{DDfock}.
Instead, they derive the DD in a covariant fashion from the handbag diagram with point-like couplings. 
By taking parametric derivatives with respect to the initial- and final-state mass-squared, 
Eq.~\eqref{DD} is derived for this model. This is the justification presented that the procedure outlined in section \ref{pion}
leads to the correct DD. Below we investigate positivity in this model and show that violation of the constraint
comes about because of the parametric differentiation.

Using the one-loop handbag diagram with scalar currents in the Bjorken limit, the vector-current DD is derived \cite{Mukherjee:2002gb}
\begin{multline} \label{FV}
F_V(x,y;t,p_1^2,p_2^2) = x \Big\{ -y(1-x-y)t \\
- x[(1-x-y)p_1^2 + y p_2^2] + m^2 \Big\}^{-1},
\end{multline}
where $p_1^2$ and $p_2^2$ are the initial- and final-state pion four-momentum squared, respectively. Here they are treated as
free parameters. Evaluating at the pion mass $M^2$, we arrive at the DD
\begin{equation} \label{F1V}
F^{(1)}_V(x,y;t) = \frac{x}{m^2 - x(1-x)M^2 - y(1-x-y)t}
\end{equation}
with a corresponding quark distribution function
\begin{equation} \label{q1V}
q^{(1)}_V(x) = \frac{x(1-x)}{m^2 - x(1-x) M^2}.
\end{equation}

To derive the DD encountered in section \ref{pion}, one applies $(p_1^2 \partial/\partial p_1^2)(p_2^2 \partial/\partial p_2^2)$ to 
$F_V$ in Eq.~\eqref{FV}. Evaluating at the pion mass, they find \cite{Mukherjee:2002gb}
\begin{equation} \label{F2V}
F^{(2)}_V(x,y;t) = \frac{2 y (1-x-y) x^3 M^4}{[m^2 - x(1-x) M^2 - y(1-x-y) t]^3},
\end{equation}
with a corresponding quark distribution
\begin{equation} \label{q2V}
q^{(2)}_V(x) = \frac{M^4}{3} \frac{x^3(1-x)^3}{[m^2 - x(1-x) M^2]^3}.
\end{equation}
Eq.~\eqref{F2V} can be manipulated into the form of Eq.~\eqref{DD} (up to the overall normalization) 
using the replacement $\lambda^2 = - M^2 /4 + m^2$. Thus one has a covariant way to derive 
power-law DDs. 

To investigate positivity, we must first derive the proper constraint. Neither models satisfy Eq.~\eqref{posi}, which is surprising
for $F^{(1)}_V$ since with point-like quark-pion coupling, there are no excluded non-two-body diagrams in the calculation. 
Eq.~\eqref{posi} holds for a composite scalar of spin-$\frac{1}{2}$ constituents. 
The constraint for scalar constituents is different as we shall now see.

For a composite scalar of scalars, a matrix element definition\footnote{We thank M.~Diehl for helping to straighten out this definition.}
of $\mathcal{F}_\zeta(X,t)$ is
\begin{equation}
\mathcal{F}_\zeta(X,t) = \frac{(2X - \zeta)  P^+}{2-\zeta} \int \frac{dz^-}{2\pi} e^{i X P^+ z^-} \langle P^\prime | q(0) q(z^-) | P \rangle
\end{equation} 
because it properly reduces to the quark distribution function
\begin{equation}
\mathcal{F}_0(X,0) = X P^+ \int \frac{dz^-}{2 \pi} e^{i X P^+ z^-} \langle  P | q(0)  q(z^-) | P  \rangle = q(X) \notag	
\end{equation}
and integrates to the form factor
\begin{equation} \label{formsum}
\int dX \; \mathcal{ F }_\zeta ( X, t) = \frac{\langle  P^\prime  | q(0) i \overset{ \leftrightarrow }\partial{}^+ q(0) | P \rangle   }{ P^+ ( 2 - \zeta) }  = F(t). 
\end{equation}

Similar to \cite{Pire:1998nw}, we can derive the relevant constraint by using translation
$q(z^-) = e^{i \mathcal{P}^+ z^-} q(0) e^{-i \mathcal{P}^+ z^-}$ and writing the Cauchy-Schwartz inequality in the following 
form
\begin{multline}
P^+ \sum_{\mathcal{Z}} \Bigg| \sqrt{X} \; \langle P | q(0) | \mathcal{Z} \rangle - a \; \sqrt{\frac{X - \zeta}{1-\zeta}} 
\; \langle P^\prime | q(0) | \mathcal{Z} \rangle \Bigg|^2 \\
\times  \delta(X P^+ + p^+_\mathcal{Z} - P^+) \geq 0
\end{multline}
for the complete set of states $\{|\mathcal{Z}\rangle\}$. Minimization with respect to the parameter $a$ yields the constraint
\begin{equation} \label{posic}
R^\prime(X,\zeta) \equiv \frac{2-\zeta}{2 X - \zeta} \sqrt{\frac{X(X - \zeta)}{q\big(X\big) q\big(\frac{X - \zeta}{1 - \zeta}\big)}}  
\; |\; \mathcal{F}_\zeta(X,0)\; | \leq 1.
\end{equation}

\begin{figure}
\begin{center}
\epsfig{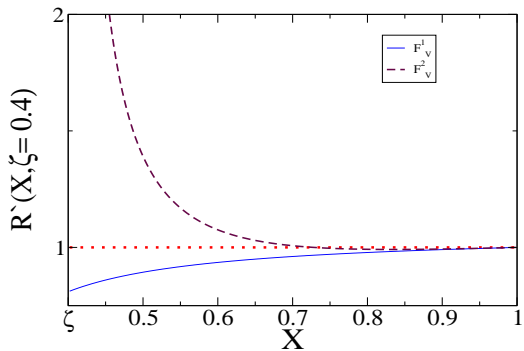}
\caption{Comparison of the scalar model, DD-based GPDs: Eq.~\eqref{F1V} ($F^1_V$) with Eq.~\eqref{F2V} ($F^2_V$)
for fixed $\zeta = 0.4$ at $t = 0$. The model parameters are chosen to be $m = 0.33$ GeV and $M = 0.14$ GeV. Here we plot the ratio $R^\prime(X,\zeta)$
appearing in equation \eqref{posic} as a function of $X > \zeta$. Positivity constrains this ratio to be less than one.}
\label{fsix}
\end{center}
\end{figure}

We can use Eq.~\eqref{posic} to test whether GPDs constructed from DDs Eqs.~\eqref{F1V} and \eqref{F2V} satisfy positivity. 
To construct their respective GPDs, we use equation \ref{gpd}. In Figure \ref{fsix} we compare the two GPDs. 
Not surprisingly $F_V^{(1)}$ satisfies the positivity constraint. As we commented above, there are no missing 
time-orderings in the calculation of this GPD since the point-like coupling between the two quarks and pion
is covariant. 

As we see from the Figure, the parametrically differentiated DD $F_V^{(2)}$ violates positivity
for all $X > \zeta$. When one treats the quarks as spin-$\frac{1}{2}$ particles, 
it is thus naturally expected that the GPD calculated from the DD Eq.~\eqref{DD} violates 
Eq.~\eqref{posi}. Viewed in this way, however, it is more of an accident that the GPD calculated 
from Eq.~\eqref{gpdreal} satisfies the constraint.


\begin{thebibliography}{99}
\bibitem{Muller:1998fv}
D.~M\"uller, D.~Robaschik, B.~Geyer, F.~M.~Dittes and J.~Ho\v{r}ej\v{s}i,
Fortsch.\ Phys.\  {\bf 42}, 101 (1994).

\bibitem{Ji:1996ek}
X.-D.~Ji,
Phys.\ Rev.\ Lett.\  {\bf 78}, 610 (1997);
Phys.\ Rev.\ D {\bf 55}, 7114 (1997).

\bibitem{Radyushkin:1996nd}
A.~V.~Radyushkin,
Phys.\ Lett.\ B {\bf 380}, 417 (1996);
{\bf 385}, 333 (1996).

\bibitem{Ji:1998pc}
X.-D.~Ji,
J.\ Phys.\ G {\bf 24}, 1181 (1998);
A.~V.~Radyushkin,
hep-ph/0101225;
K.~Goeke, M.~V.~Polyakov and M.~Vanderhaeghen,
Prog.\ Part.\ Nucl.\ Phys.\  {\bf 47}, 401 (2001).

\bibitem{Mukherjee:2002gb}
A.~Mukherjee, I.~V.~Musatov, H.~C.~Pauli and A.~V.~Radyushkin,
hep-ph/0205315.

\bibitem{Radyushkin:1997ki}
A.~V.~Radyushkin,
Phys.\ Rev.\ D {\bf 56}, 5524 (1997).

\bibitem{Radyushkin:1998es}
A.~V.~Radyushkin,
Phys.\ Rev.\ D {\bf 59}, 014030 (1999).

\bibitem{Frederico:2001qy}
T.~Frederico and H.~C.~Pauli,
Phys.\ Rev.\ D {\bf 64}, 054007 (2001).

\bibitem{Drell:1969km}
S.~D.~Drell and T.-M.~Yan,
Phys.\ Rev.\ Lett.\  {\bf 24}, 181 (1970);
G.~B.~West,
Phys.\ Rev.\ Lett.\  {\bf 24}, 1206 (1970).

\bibitem{Mankiewicz:1997uy}
L.~Mankiewicz, G.~Piller and T.~Weigl,
Eur.\ Phys.\ J.\ C {\bf 5}, 119 (1998).

\bibitem{Brodsky:2001xy}
S.~J.~Brodsky, M.~Diehl and D.~S.~Hwang,
Nucl.\ Phys.\ B {\bf 596}, 99 (2001);

\bibitem{Diehl:2000xz}
M.~Diehl, T.~Feldmann, R.~Jakob and P.~Kroll,
Nucl.\ Phys.\ B {\bf 596}, 33 (2001)
[Erratum-ibid.\ B {\bf 605}, 647 (2001)].


\bibitem{Saull:1999kt}
P.~R.~Saull  [ZEUS Collaboration],
hep-ex/0003030;
A.~Airapetian {\it et al.}  [HERMES Collaboration],
Phys.\ Rev.\ Lett.\  {\bf 87}, 182001 (2001);
C.~Adloff {\it et al.}  [H1 Collaboration],
Phys.\ Lett.\ B {\bf 517}, 47 (2001);
S.~Stepanyan {\it et al.}  [CLAS Collaboration],
Phys.\ Rev.\ Lett.\  {\bf 87}, 182002 (2001).

\bibitem{Diehl:1997bu}
M.~Diehl, T.~Gousset, B.~Pire and J.~P.~Ralston,
Phys.\ Lett.\ B {\bf 411}, 193 (1997).

\bibitem{Belitsky:2000gz}
A.~V.~Belitsky, D.~M\"uller, L.~Niedermeier and A.~Sch\"afer,
Nucl.\ Phys.\ B {\bf 593}, 289 (2001).

\bibitem{Petrov:1998kf}
V.~Y.~Petrov, P.~V.~Pobylitsa, M.~V.~Polyakov, I.~B\"ornig, K.~Goeke and C.~Weiss,
Phys.\ Rev.\ D {\bf 57}, 4325 (1998).

\bibitem{Guidal:2002kt}
M.~Guidal and M.~Vanderhaeghen,
hep-ph/0208275.

\bibitem{Pire:1998nw}
B.~Pire, J.~Soffer and O.~Teryaev,
Eur.\ Phys.\ J.\ C {\bf 8}, 103 (1999).

\bibitem{Pobylitsa:2001nt}
P.~V.~Pobylitsa,
Phys.\ Rev.\ D {\bf 65}, 077504 (2002);
{\bf 65}, 114015 (2002).


\bibitem{Tiburzi:2001ta}
B.~C.~Tiburzi and G.~A.~Miller,
Phys.\ Rev.\ C {\bf 64}, 065204 (2001).

\bibitem{Tiburzi:2001je}
B.~C.~Tiburzi and G.~A.~Miller,
Phys.\ Rev.\ D {\bf 65}, 074009 (2002).

\bibitem{Tiburzi:2002mn}
B.~C.~Tiburzi and G.~A.~Miller,
hep-ph/0205109.


\end{thebibliography}
\end{document}